\documentclass[
   ,final            
  ]
  {aipproc}

\layoutstyle{8x11single}

\usepackage{amsmath,amssymb}
\usepackage[usenames]{color}

\newcommand{\EF}[1]{}
\newcommand{\AK}[1]{}
\newcommand{\MP}[1]{}  

\begin{document}

\title{Parallel kinetic Monte Carlo simulation
of Coulomb glasses}

\classification{02.70.Tt, 71.23.Cq, 72.20.Ee}

\keywords      {Monte Carlo algorithms, Hopping transport}

\author{E.E.~Ferrero}{
  address={CONICET, Centro At\'omico Bariloche, 8400 Bariloche, Argentina}
  , altaddress={Current address: LIPhy, Universit\'e Joseph Fourier, UMR 5588, F-38402 Saint Martin d'H\`eres, France} 
}

\author{A.B. Kolton}{
  address={CONICET, Centro At\'omico Bariloche, 8400 Bariloche, Argentina}
}

\author{M. Palassini}{
  address={Departament de F\'{\i}sica Fonamental, Universitat de Barcelona, 08028 Barcelona, Spain},altaddress={The Abdus Salam International Centre for Theoretical Physics (ICTP), 34014 Trieste, Italy} 
}

\begin{abstract}
We develop a parallel rejection 
algorithm to tackle the problem of low acceptance in
Monte Carlo methods, and apply it to the simulation 
of the hopping conduction in Coulomb glasses using
Graphics Processing Units, for which we also parallelize
the update of local energies. In two dimensions, our parallel code 
achieves speedups of up to two orders
of magnitude in computing time over an equivalent serial code.
We find  numerical evidence of a scaling relation
for the relaxation of the conductivity at different temperatures.

\end{abstract}

\maketitle


\section{Introduction}
A frequent limitation of Markov-chain Monte Carlo (MC) methods is low acceptance. In {\em equilibrium} MC this problem can sometimes be circumvented
with a clever choice of the MC moves, which can be chosen with a certain freedom 
as long as the Markov chain
converges to the probability distribution of interest.
Such freedom is not allowed in {\em kinetic} Monte Carlo (KMC), in which the MC moves
are dictated by the physical dynamics to be simulated.

In this work we present a general ``parallel rejection'' (PR) algorithm to 
address the problem of low acceptance, which 
 is especially suitable for implementation on Graphics Processing Units 
(GPUs), easily available and inexpensive platforms for 
massively parallel computing.
We apply PR to the KMC simulation of the hopping conduction
in Coulomb glasses, which is notoriously plagued by very low acceptance 
in the variable-range hopping (VRH) regime.
For this particular application, PR can be seen 
as a parallelization of the ``mixed'' algorithm
of Refs.~\cite{tsigankov1,tsigankov2}, which 
can be further accelerated in GPUs 
by efficiently parallelizing the $N$ local energy updates 
that, due to the long-range interaction, are required after 
each elementary MC move in a system with $N$ sites.

We implemented a GPU code using CUDA~\cite{CUDA}, exploiting
both sources of parallelism.
In the next section we illustrate the PR idea in general, and later
we apply it to the hopping dynamics of the Coulomb glass.
We then present our results for the lattice
model in two dimensions,
notably on the short-time relaxation of the conductivity.
 Finally, we compare the performance of the GPU code with a serial implementation of the mixed algorithm.

\section{Parallel rejection algorithm}
Let us consider a generic Markov chain
specified by a proposal matrix $Q_{\alpha \beta}$ and 
an acceptance matrix $P_{\alpha \beta}$ between
the configurations $\alpha, \beta$ of a certain system.
The standard (serial) {\em rejection algorithm} to simulate such a chain 
consists in iterating the following two steps:
1. Propose a move $\alpha \to\beta$, where $\alpha$ is the
current configuration and $\beta$ is chosen with probability 
$Q_{\alpha \beta}$; 
2. Accept the move with probability 
$P_{\alpha \beta}$ and, if this is accepted, update the configuration to $\beta$.

The PR algorithm (see Ref.\cite{niemi} for a similar idea)
runs simultaneously on an ordered 
array of $M$ parallel ``threads'' (for example, GPU threads),
iterating the following steps ($k=1,\dots, M$ is the thread label):
\begin{itemize}
\item[1.] 
Propose, independently from the other threads,
a move $\alpha \to \beta_k$, where 
$\alpha$ is the
current configuration (common to all threads) and $\beta_k$ 
is chosen with probability $Q_{\alpha \beta_k}$.

\item[2.] 
Accept the move $\alpha \to \beta_k$ with probability $P_{\alpha \beta_k}$, 
independently from the other threads (without updating the configuration).

\item[3.] If at least one thread has accepted a move, 
update the configuration 
to $\beta_{q}$, where $q$ is the lowest label among the threads that have 
accepted a move.
\end{itemize}

\begin{figure}[tt]
  \label{fig:parallelrejection}
  \includegraphics[scale=0.25]{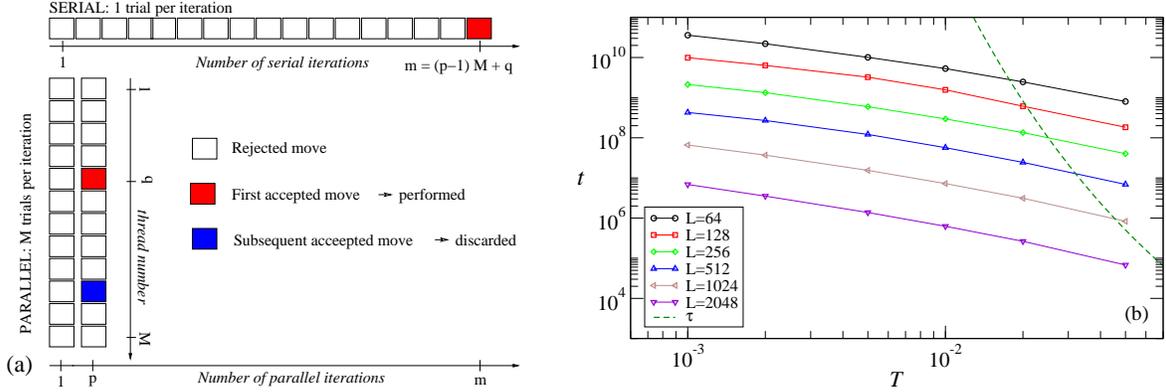}
  \hspace{0.5cm}
  \includegraphics[scale=0.33]{figs/maxwelltime.eps}
  \caption{\small
  (a) Illustrative comparison of the ``Parallel Rejection'' scheme with  
  the mathematically equivalent serial approach. 
  In both, proposed moves (squares) are drawn by sampling from the proposal probability.  
  While in the serial approach (top) one move is proposed each iteration, in 
  the parallel approach (bottom) a block of $M$ independent moves is proposed. 
  For a given acceptance rate, the first accepted move (red square) is found 
  after $m$ serial iterations, and after $p$ parallel iterations. 
  Other accepted moves (blue square) that may appear in last parallel iteration are discarded.
In this example  $M=12$, $p=2$, $q=5$. 
  (b) Physical time (in units of $\tau_0$) that we can simulate with
 the parallel KMC method in a 10 days run, as a function of temperature, 
  for different systems sizes.
  The dashed line is the  characteristic  time 
  $\tau = b^{-1} T^{-1} \exp[(T_0/T)^{1/2}]$, with $b^{-1}=2.7, T_0=6.2$
  according to 
  Fig.\ref{fig:conductance}.
  } 
\end{figure}

While the two algorithms above are mathematically equivalent
(see Fig.\ref{fig:parallelrejection}(a) for an illustration), the
parallel version is increasingly faster as  
the acceptance rate, $A=\langle P_{\alpha \beta} \rangle$, decreases.
If $m$ and $p$ are the number of iterations of the serial and parallel
algorithms, respectively, until a move is accepted, 
we can estimate the speedup of PR as $\langle m \rangle/\langle p\rangle$
(neglecting parallelization overheads and differences in the computing time
for one iteration in the parallel and serial implementations).
Since $\langle m \rangle = A^{-1}$ and $\langle p\rangle=
1/[1-(1-A)^M]$, for $A\ll 1$ we have 
$\langle p \rangle \sim (M A)^{-1} \ll \langle m \rangle$, and thus 
 $\langle m \rangle/\langle p\rangle \sim {\mbox{min}}(A^{-1},M)$.
In principle, the optimal choice for $M$ is thus $M^* = {\mbox{min}}(A^{-1},C)$,
where $C$ is the maximum number of threads that can run simultaneously.
In practice, in a GPU implementation it is possible (and recommended)
to use $M$ larger than $C$, in which case
the virtual parallel execution is efficiently handled by the hardware scheduling system.
The computing time can then remain sublinear in $M$ even for $M$ as large as
$10^4$. Thus, for many applications the speedup of the PR algorithm on GPUs is 
essentially dictated by $~A^{-1}$, and can be very significant.

A popular alternative to the rejection algorithm in low-acceptance 
situations is the rejection-free BKL or Gillespie algorithm 
\cite{BKL}.
This requires a computing time proportional to the average
number $z=\langle \sum_\beta \theta(Q_{\alpha\beta}P_{\alpha \beta})\rangle$
of configurations that are accessible from a given configuration $\alpha$.
Thus, BKL is faster than serial rejection when $z \ll A^{-1}$,
but slower than PR when $(M A)^{-1} \ll z$.
Below we apply PR to the rejection part of the mixed algorithm 
of Ref.~\cite{tsigankov1}, which combines the BKL and rejection 
algorithms for the simulation of phonon-assisted hopping conduction.

\section{Hopping conduction in the Coulomb glass}

We consider the standard Coulomb glass model~\cite{Efros} described by
the dimensionless Hamiltonian
\begin{equation}
  H = \sum \phi_i n_i + \frac{1}{2}\sum_{i\neq j} \frac{(n_i-\nu)(n_j-\nu)}{r_{ij}}-E \sum_i n_i x_i,
\end{equation}
where $n_i=0,1$ ($i=1,\dots, N)$, with $\sum_i n_i = \nu N$,
 are the electron occupation numbers of $N$ sites in a $D$-dimensional 
volume, 
 $\nu$ is the filling factor,
$r_{ij}$ is the Euclidean distance between sites $i$ and $j$
in units of the average spacing $a=({\mbox{Volume}}/N)^{1/D}$, and $E$ is an external electric field in
the (negative) $x$ direction. 
The random potentials $\phi_i$ are sampled from the uniform
distribution $[-W,W]$.
We consider only phonon-assisted single-electron hops, with transition
rates that can be approximated by
\begin{equation}
  \Gamma_{i j} = \tau_0^{-1} \theta_{ij} e^{-2r_{ij}/\xi} \min[1,e^{-\Delta H_{ij}/ k_B T}]
\label{eq:gamma}
\end{equation}
where $T$ is the temperature, $k_B$ is the Boltzmann constant, 
$\theta_{ij}=\delta_{n_i,1}\delta_{n_j,0}$, 
$\xi$ is the electron localization length, and $\tau_0$ is a microscopic
time of order $10^{-12}$s, which will be our unit of time.
The energy change after a hop $i\to j$ is 
$\Delta H_{ij} = \epsilon_j -\epsilon_i - 1/{r_{ij}} - E \Delta x_{ij},$ 
where $\epsilon_i = \phi_i  + \sum_j (n_j-\nu)/r_{ij}$ are the
single-particle energies and
$\Delta x_{ij}=x_j-x_i$ is the length of the hop along the field.
The instantaneous conductivity is given by \cite{tsigankov1,tsigankov2}
\begin{equation}
  \sigma(t) = \frac{1}{N E} \frac{dP(t)}{dt}
\label{sigmadef}
\end{equation}
where $P = \sum_i n_i x_i$ is the electric dipole moment, provided
$E$ is small enough to ensure a linear response.

In the naive serial KMC algorithm for simulating the dynamics in 
Eq.(\ref{eq:gamma}) a hop $i\to j$ is proposed by choosing $i$ and $j$ 
uniformly at random among the $N$ sites, and is accepted
with probability $\tau_0 \Gamma_{ij}$  (thus, $Q_{\alpha \beta}=
N^{-1} (N-1)^{-1}\sum_{ij}\Theta^{ij}_{\alpha\beta}$ and $P_{\alpha \beta}=
\sum_{ij}\tau_0 \Theta^{ij}_{\alpha\beta} \Gamma_{ij}$, 
where $\Theta^{ij}_{\alpha\beta} =1$ if $\alpha$ and $\beta$ differ
by the hop $i\to j$ and $\Theta^{ij}_{\alpha\beta}=0$ otherwise).
The time is incremented by $\Delta t=1$
after $N (N-1)$ proposals. 

The above algorithm
suffers from extremely low acceptance due to both the tunneling factor
$\Gamma_{ij}^T = e^{-2r_{ij}/\xi}$ and the thermal activation factor
$\Gamma_{ij}^A = \theta_{ij}\min[1, \exp(-\Delta H_{ij}/T)]$.
The mixed algorithm \cite{tsigankov2} exploits the factorization
$\tau_0 \Gamma_{ij}=\Gamma^T_{ij} \Gamma^A_{ij}$ and the fact that
$\Gamma^T_{ij}$ is configuration independent. 
The proposal matrix is now 
$Q_{\alpha \beta}=
\sum_{ij}\Theta^{ij}_{\alpha\beta} \Gamma^T_{ij}/\Gamma$,
where  $\Gamma  = \sum_i \sum_{j\neq i} 
\Gamma^T_{ij}$, and can be sampled without rejection (for example,
using the ``tower sampling'' method ~\cite{towersampling}).
The acceptance matrix is
$P_{\alpha \beta}=\sum_{ij}\Theta^{ij}_{\alpha\beta} \Gamma^A_{ij}$.
After each proposal, $t$ is incremented by
a random  $\Delta t$ sampled from $p(\Delta t) = \Gamma \exp(-\Gamma \Delta t)$
\cite{BKL}. Since the rejection now only comes from $\Gamma^A_{ij}$, 
the acceptance rate is increased by a factor $\sim N \xi^{-D}$. Nevertheless, 
deep in the VRH regime, where $T$ is only a few 
percent of the Coulomb energy, 
the acceptance is still quite low 
(for the $D=2$ lattice model we 
find $A \approx 0.03 \,T$ for $T\leq 0.2$).
It becomes then advantageous to use the PR strategy.
This results in the following mixed PR algorithm running on $M$ threads ($k=1,\dots,M$), which is mathematically equivalent to the serial
mixed algorithm and, therefore, to the naive KMC:

\begin{itemize}
\item[1.] Propose (indendently from the other threads) a 
hop $i_k \to j_k$ by choosing $(i_k,j_k)$ 
with probability $\Gamma^T_{i_k j_k}/\Gamma$ by rejection-free sampling.

\item[2.] Accept  
the hop with probability $\Gamma^A_{i_kj_k}$ 
(even if accepted, do not execute the hop).

\item[3.] If at least one thread has accepted a hop then:

\begin{itemize}
\item[a.]
Execute the hop $i_q\to j_q$,
where $q$ is the lowest label among the threads that have 
accepted a hop. Any accepted hop in the other threads is discarded. 

\item[b.] Update the dipole moment as $\Delta P= \Delta x_{i_q j_q}$ and the 
local energies by adding $1/r_{i_q j_q}$ to $\epsilon_{i_q}$,
$-1/r_{i_q j_q}$ to $\epsilon_{j_q}$, and 
$1/{r_{lj_q}}-1/{r_{li_q}}$ to $\epsilon_l$ for $l\neq i_q, l \neq j_q$.

\item[c.] Increment the time by a random $\Delta t$ 
sampled from a Gaussian  
distribution of mean $m/\Gamma$ and standard deviation $\sqrt{m}/\Gamma$, 
where $m = p M + q$ and $p$ is the number
of iterations since the previous executed hop.
\end{itemize}
\end{itemize}

\begin{figure}[tt]
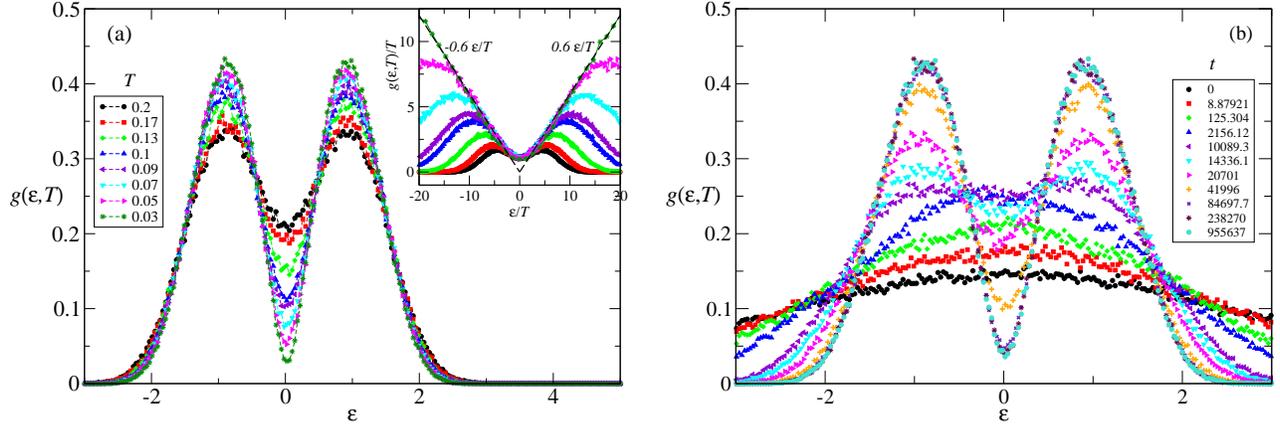

  \label{fig:dos}
  \includegraphics[scale=0.33]{figs/DOSpsequil_vsT.eps}
  \hspace{0.3cm}
  \includegraphics[scale=0.33]{figs/DOSnoneq_vs_time} 
  \caption{Pseudo-equilibrium (a) and time-dependent (b) single-particle density
of states for $L=512$, as function of
  temperature and time respectively. Data in (b) corresponds to $T=0.04$.}
\end{figure}

We implemented the above algorithm in CUDA~\cite{CUDA}
for the case in which the sites belong to a lattice with toroidal 
boundary conditions.
In this case $\Gamma^T_{ij}$ only depends on $r_{ij}$ 
(thus $\Gamma = N \sum_i \Gamma^T_{ij}$) hence 
the ``tower'' is much smaller than for the model with random sites 
\cite{tsigankov2}. At each iteration we need 
$3 M$ random numbers (to choose $i_k$, $r_{i_k j_k}$, and to accept the hops), 
which requires a random number generator (RNG) 
able to generate a large number of uncorrelated random sequences in parallel.
The code uses functions of our own and 
the open-source libraries 
Thrust~\cite{THRUST} (for generic parallel transformations) 
and Philox from Random123~\cite{PHILOX} (a GPU-suitable RNG).
The code also exploits the ``embarrassing parallelism'' of the 
local energy update:
we distribute the update on $N$ parallel threads,
where the $k$-th thread updates the local energy on lattice site $k$.

\section{Results for the lattice model in  two dimensions}

We simulated two-dimensional square lattices 
with $N=L^2$ sites, for $L=64,128,256,512,1024,2048$,
setting $\nu=1/2$, $\xi=1$, $W=1$, and $E=T/10$.
We use toroidal boundary conditions in both directions, 
and adopt the minimal image convention for $r_{ij}$ 
and $\Delta x_{ij}$.
We do not allow hops larger than $L/2$ (in our simulations,
the typical hopping  length is much shorter than $L/2$ anyway).
To recover cgs units from the numerical data below, the 
dimensionless quantities $\{H, T, t, E, P, \sigma\}$ must
be multiplied respectively by $\{e^2/(\kappa a), e^2/(\kappa a k_B), \tau_0, e/(\kappa a^2), 
e a, a^{3-D} \kappa/\tau_0\}$, where $e$ is the electron charge
and $\kappa$ is the dielectric constant of the lattice.

To validate the code, we start by analyzing the single-particle density of states, defined as
the normalized histogram $g(\epsilon,T)=\sum_k \delta_{\Delta \epsilon}(\epsilon - \epsilon_k)$, 
for a given binning size $\Delta \epsilon$.
In Fig.\ref{fig:dos}(a) we show $g(\epsilon,T)$ in the steady 
state for different temperatures and $L=512$. As shown in the inset, 
in the Coulomb gap region the data can be rescaled as $g(\epsilon,T)
=T f(\epsilon/T)$, and are well fitted by $g(\epsilon,T) = c |\epsilon|$
for small $|\epsilon|$, with $c=0.62$.
The prefactor is consistent with the theoretical estimate~\cite{Efros} $c=2/\pi$
but is larger than the estimate $c=0.40$ obtained 
with the parallel tempering MC algorithm \cite{goethe_palassini},
which allows many-electron rearrangements and thus can reach 
lower energies then the single-particle KMC, even if
the latter seems to have reached a steady state, as shown in 
Fig.\ref{fig:dos}(b)).

\begin{figure}[tt]
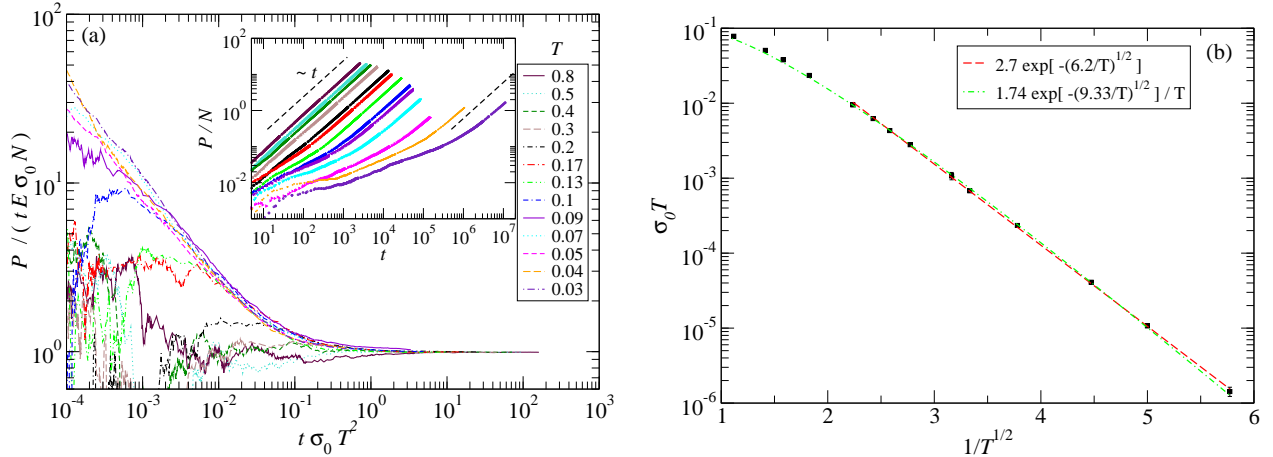

  \label{fig:conductance}
  \includegraphics[scale=0.33]{figs/Pvst}
  \hspace{0.3cm}
  \includegraphics[scale=0.33]{figs/ESlaw2} 
  \caption{\small
    (a) Inset: time evolution of the polarization starting from a random
    initial configuration at different temperatures for $L=512$.
    The steady-state conductivity $\sigma_0 = \lim_{t\to \infty} (dP/dt)/(N E)$ is reached after a transient 
    time $\tau \approx 1/(T^2 \sigma_0)$.
    Main figure: same data rescaled as a function of $t/\tau$. 
In this figure we use a deterministic time increment 
$\Delta t=m/\Gamma$. Using a stochastic increment does not produce
any appreciable difference.
    (b) Fit 
to the Efros-Shklovskii law $\sigma_0 = b T^{-\lambda} \exp[-(T_0/T)^{1/2}]$ for $\lambda=1, 2$. 
    }
\end{figure}

Next, we analyze the conductivity.
The inset of Fig.\ref{fig:conductance}(a) shows how, starting
from a random configuration, after a transient time
the polarization grows linearly in time, and a stationary 
conductivity can be estimated as $\sigma_0 = P/(N E t)$ for large $t$.
Further relaxation of the conductivity at larger times cannot be
discarded, but should be neglibible 
for time scales a few times larger than the transient time.
Our data for $\sigma_0(T)$, shown in Fig.\ref{fig:conductance}(b), agree
 with Ref.\cite{tsigankov1} up to a factor two.
The data are well fitted by the Efros-Shklovskii law
$\sigma_0 = b T^{-\lambda} \exp[-(T_0/T)^{1/2}]$ assuming $\lambda=1$,
which gives $T_0=6.2, b=2.7$. 
However, the choice $\lambda=2$ fits equally well the data for $T<0.25$,
giving $T_0=9.3, b=1.7$. 
As shown in
the main figure, the time evolution of $P(t)$ at different temperatures can be collapsed
very well onto a single curve by rescaling the time
with a characteristic transient time $\tau \sim (\sigma_0 T^2)^{-1}$.
This suggests a scaling relation
\begin{equation}
\sigma(t, T) = \sigma_0(T) f(t \sigma_o(T) T^2) \,.
\label{eq:scalingsigma}
\end{equation}
We checked that finite-size effects on $P(t)$ are negligible.
The proportionality of $\tau$ to $\sigma_0^{-1}$ is to be expected
\cite{tsigankov2,bergli-galperin}, since the typical hopping time
of the current-carrying hops is $\sim \exp[(T_0/T)^{1/2}]$. 
A discussion of the significance of the factor $T^2$ 
(both in $\tau$ and, possibly, in the Efros-Shklovskii law) is outside
the scope of this paper, but we note that
using the scaling variable $t/(\sigma_0 T)$ we obtain a much worse data
collapse (not shown). It is interesting to compare Eq.(\ref{eq:scalingsigma})
with the scaling $\sigma(\omega,T) = \sigma_0(T) f(\omega/T\sigma_0(T))$
for the low-frequency ac conductivity found in Ref.\cite{bergli-galperin2}.

\section{Code performance}
\label{sec:performance}

In this section we compare the performance of our parallel GPU code with 
that of a serial CPU implementation of the mixed algorithm.
In both cases, the wall-clock computing time required to execute a hop
is the sum of two main contributions: 
the {\em rejection time}, $W_R$, spent proposing and rejecting hops until one is
accepted, and the {\em update time}, $W_U$, spent updating the $N$ local energies after a
hop is executed.
As discussed earlier, aside from hardware-related corrections  
we expect 
$W_R \propto A^{-1} \log L$ for the CPU code and $W_R \propto (A M)^{-1} \log L$ 
for the GPU code running on $M$ threads,
 where the acceptance $A$ increases 
with the temperature and the $\log L$ factor comes from the 
tower-sampling.
$W_U$ is independent of the temperature and the configuration, 
and we expect $W_U \propto L^2$.
In the parallel implementation, nevertheless, this will hold
only at large enough $L$ (when the hardware occupancy saturates), while at smaller sizes 
the scaling will be sublinear in $L^2$, and even constant for very small sizes.
In the Coulomb glass, most excitations active at low $T$ are {\em dipoles}
(short electron-hole pairs). In 2D the density of states of dipoles 
at low energy is constant, which implies $A\propto T$ \textit{in the steady state} (indeed we find $A \approx 0.03 \,T$). 
Hence, the overall computing time will be dominated by $W_R$
for temperatures below a certain threshold that decreases with $L$ as $\log L/L^2$, 
and by $W_U$ above the threshold.

\begin{figure}[tt]
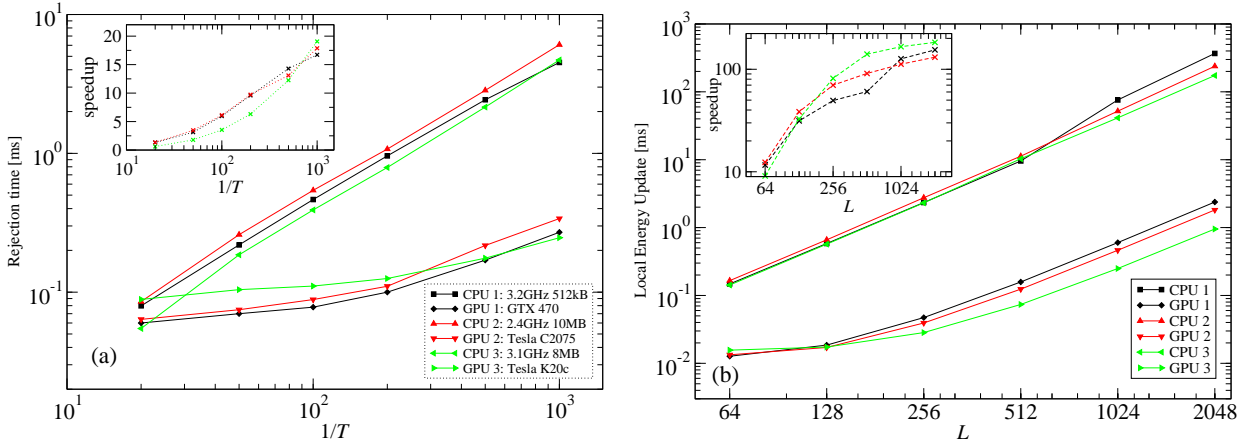

  \includegraphics[scale=0.33]{figs/fig-performance-a.eps}
  \hspace{0.2cm}
  \includegraphics[scale=0.33]{figs/fig-performance-b.eps}
  \caption{\small
Computing time [ms] per executed 
hop for the serial and parallel codes in three different CPU+GPU  platforms.
  We benchmark a fair serial single CPU core implementation 
  against our parallel GPU implementation (with the same CPU core as host),
both with double precision floating point operations.
  (a) Temperature dependence of the average rejection time ($W_R$)
for $L=256$.
(b) Size dependence of the average update time ($W_U$) of the $N=L^2$ local energies.
  The insets show the speedup of the GPU over the CPU implementation.
  While the speedup starts to saturate with $L$ in (b), 
it is still strongly growing with $1/T$ in (a).
  }
\label{fig:separateperformances}
\end{figure}

To see how well the above estimates hold in practice,
in Fig.\ref{fig:separateperformances}(a) we show
the temperature dependence of $W_R$ for the CPU and GPU
codes in the steady state for $L=256$ (the dependence on $L$ is very weak).
For the CPU code, we see roughly 
$W_R \propto 1/T$, as expected.
For the GPU code we used a number of threads 
close to the optimal value $M^* \approx A^{-1}$.
Hence, at moderate temperatures where $M^*$ is not too large,
$W_R \propto (M^* A)^{-1}$ is almost independent of $T$, as expected. 
At very low $T$, $M^*$ is large enough to saturate the maximum number of 
concurrent threads and thus $W_R$ increases with $1/T$, although less 
than linearly since the GPU can still save time by hiding memory latency.
Therefore, the relative performance of the GPU vs 
the CPU codes, shown in the inset, is consistent with the expected
linear behavior in $1/T$. 

In Fig.\ref{fig:separateperformances}(b) we compare the size dependence
of $W_U$ for the GPU and CPU implementations.
As expected, for the CPU we observe $W_U\propto L^2$ while 
for the GPU, using $L^2$ threads, $W_U$ is almost constant for small $L$ 
and grows for large $L$. Interestingly, the increase is still sublinear even 
when $L^2$ is several times larger than the 
number of physical cores, due to the efficient internal thread scheduling.
Consequently the speedup, which is already substantial 
even for relatively small sizes, increases with $L$ and exceeds
two orders of magnitude at $L=2048$, without quite saturating yet.

The overall computing time \textit{per executed hop} of 
the GPU code is basically the sum of $W_R$ and $W_U$.
For example, for the platform
CPU3+GPU3 (see Fig.4 for details) the overall time ranges from 0.1 ms ($L$=64) to 1.04 ms ($L$=2048)
at $T=0.05$, representing speedups of 1.7x to 179x respect to the serial code, 
and 0.27 ms ($L$=64) to 1.23 ms ($L$=2048) at $T=0.001$, representing
speedups of 15x to 157x. 
Nevertheless, note that these speedup factors depend on the particular 
serial implementation and hardware.

In order to see in what regime of $T$ and $L$ the GPU code may 
be useful in practice,
it is illustrative to estimate the physical time we can simulate in, say, a ten-day
run. While this time is independent of $T$
for the CPU code, according to the previous discussion it scales 
as $1/T$ for the GPU code for low enough $T$.
This is confirmed in Fig.\ref{fig:parallelrejection}(b),
where we also show the transient time $\tau=1/(\sigma_0 T^2)$ 
for the establishment of a steady-state conductivity.
Clearly, since $\tau$ grows much faster than the speedup as $T$ decreases,
even the GPU code cannot reach the steady state at very low $T$. 
At the temperatures at which we {\em can} reach the steady state,
the typical hopping length  $r \sim (\xi/4)(T_0/T)^{1/2}$
is less than ten lattice spacings.
Hence, for the purpose of measuring the conductivity 
it is preferable to average over many 
samples at intermediate $L$, rather than a few samples at large $L$.
Large samples, for which the GPU code is significantly advantageous, 
might be useful for studying the large-scale 
geometry of the conducting paths, for instance.

It is worth noting that the GPU code should be 
significantly more advantageous in 3D than in 2D,
because
(i) $N$ increases more rapidly with $L$, and   
(ii) both $T_0$ and the dipole density of 
states of dipoles are smaller in 3D, so 
in the VRH regime the acceptance rate is even lower.
It is also straightforward to incorporate multiple-electron
hops in our GPU code. Since these hops have even lower acceptance rate,
we expect the speedup to be substantial.

\section{Conclusions}

We have presented a novel parallel KMC technique 
for simulating the Coulomb glass hopping conduction.
It allows to simulate larger systems and longer times 
than its serial counterpart, with speedups over 100x 
for relatively 
large system sizes in two dimensions.
This might be helpful for studying features involving larger 
length-scales. In 2D, we find that the short-time relaxation of the
conductivity at different temperatures is well described by a single
scaling curve.
Finally, our current implementation can be easily extended to 
higher dimensions, multiple occupation, and multi-electron hops.
For these extensions we can expect an even larger speedup 
with respect to the mathematically equivalent serial implementation.
The code is available to download, modify and use under GNU GPL 3.0 at~\cite{code}.

\begin{theacknowledgments}

We acknowledge support from the ``Acci\'on Integrada Argentina-Espa\~na'' MINCYT-MINECO 
(Ref. PRI-AIBAR-2011-1206). 
MP acknowledges support from MINECO (FIS2012-38266-C02-02) and AGAUR (2012 BE 00850).
We thank FaMAF-UNC (Argentina), and LIPhy-UJF (France) for computational resources
for benchmarking and simulations. MP thanks the CMSP section
of ICTP for hospitality.
\end{theacknowledgments}


\bibliographystyle{aipproc}   

\bibliography{ferrero_kolton_palassini}

\IfFileExists{\jobname.bbl}{}
 {\typeout{}
  \typeout{******************************************}
  \typeout{** Please run "bibtex \jobname" to optain}
  \typeout{** the bibliography and then re-run LaTeX}
  \typeout{** twice to fix the references!}
  \typeout{******************************************}
  \typeout{}
 }

\end{document}